      [1999/12/01 v1.4c Il Nuovo Cimento]
\documentclass{cimento}
\usepackage{graphicx}  
\def\ltsima{$\; \buildrel < \over \sim \;$}
\def\simlt{\lower.5ex\hbox{\ltsima}}
\def\gtsima{$\; \buildrel > \over \sim \;$}
\def\simgt{\lower.5ex\hbox{\gtsima}}
\title{Rest frame light curves of Swift GRBs}
\author{V.~Mangano\from{ins:1}\ETC,
V.~La Parola\from{ins:1}, E.~Troja\from{ins:1}, G.~Cusumano\from{ins:1}, 
T. Mineo\from{ins:1}, D.~Burrows\from{ins:2}, S.~Campana\from{ins:3}, 
M.~Capalbi\from{ins:4}, G.~Chincarini\from{ins:3}\from{ins:5}, N. Gehrels\from{ins:6},
P.~Giommi\from{ins:4}, A.~Moretti\from{ins:3}, M.~Perri\from{ins:4}, 
P.~Romano\from{ins:3}, 
\atque G.~Tagliaferri\from{ins:2}} 
\instlist{\inst{ins:1}INAF-IASF PA, Via Ugo La Malfa 153, I-90146, Palermo, Italy
\inst{ins:2}Dept of Astronomy \& Astrophysics, PSU, University Park, PA 16802, USA 
\inst{ins:3}INAF-Osservatorio Astronomico di Brera, Via E. Bianchi 46, I-23807 Merate (LC), Italy
\inst{ins:4}ASI-Science Data Center, Via G. Galilei, I-00044, Frascati (RM), Italy
\inst{ins:5} Universit\`a{} degli Studi di Milano, Bicocca, Piazza delle Scienze 3, I-20126, Milano, Italy
\inst{ins:6}NASA/Goddard Space Flight Center, Greenbelt, MD20771, USA }
\PACSes{\PACSit{98.70.Rz}{$\gamma$-ray source; $\gamma$-ray bursts}
\PACSit{01.30.Cc}{Conference proceedings}}
\begin{document}

\maketitle

\begin{abstract}
We have computed the luminosity rest frame light curves of the first 40
Gamma-ray bursts (GRBs) detected by Swift with well established redshift.
We studied average properties of the light curves in the four subsamples
of bursts given by $z<1$, $1<z<2$, $2<z\simlt4$, and $z\simgt4$. We conclude
that all the last three subsamples share the same morphology and the
same luminosity range. Very high redshift ($z\simgt4$) GRBs detected up to
now are not intrinsically longer than lower redshift long GRBs. Nearby
long GRBs ($z<1$) are fainter than average. Possible selection effect are
under investigation. 
\end{abstract}

\begin{figure}[t]
\hspace{-0.5truecm}
\includegraphics[height=7.5cm,width=12cm,angle=0]{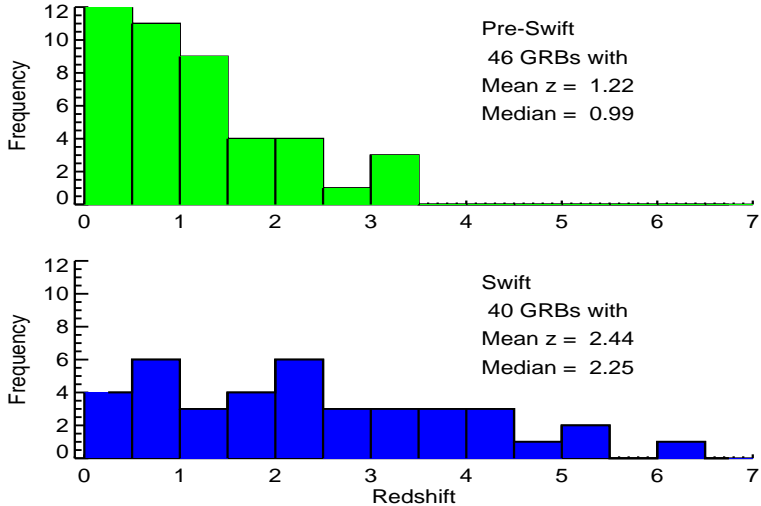} 
\caption{Redshift distributions of the pre-Swift and Swift GRBs}  
\label{redshiftdist}
\end{figure}

We have computed the 0.2$-$10 keV luminosity rest frame light curves 
of the first 40 GRBs detected by Swift with well established redshift. 
The sample covers GRBs up to May 2006 and includes of 9 very high redshift 
($z \simgt$4) GRBs
and 3 short GRBs (GRB~050509B, GRB~050724 and GRB~051221A, all with
redshift $z<1$). 
More than half of the selected sources (26 over 40) show flares 
during  the XRT observation. 
%
%
Our sample is definitely larger than samples used in previous works 
on Swift GRBs luminosity light curves 
(Chincarini {\it et al.} 2005, 7 GRBs used; Nousek {\it et al.} 2006, 14 GRBs used) 
and is comparable to the sample of GRBs with
redshift triggered by the pre-Swift missions
({\it e.g}. BeppoSAX, HETE~II, INTEGRAL) that consists of 46 objects.
However, the redshift distributions of the two samples are
substantially different (see Fig.~\ref{redshiftdist}):
pre-swift bursts have an average redshift of 1.22 and a 
larger fraction of the population at low redshift,
while Swift bursts have an average redshift of 2.44 and
a flat distribution up to $z=5$, with a single
source at redshift $z=6.3$ (GRB~050904; Cusumano {\it et al.} 2006).
Several factors can shape the redshift distribution of Swift detected 
bursts, and among them a very relevant contribution may come from
the narrowness of the BAT sensitivity range
(15$-$150~keV) that favors detection of GRBs with a peak
energy close to this range, and in particular, high redshift
GRBs, because their typical 300~keV peak energy is easily lowered 
by cosmological redshift effects.

\begin{figure}[t]
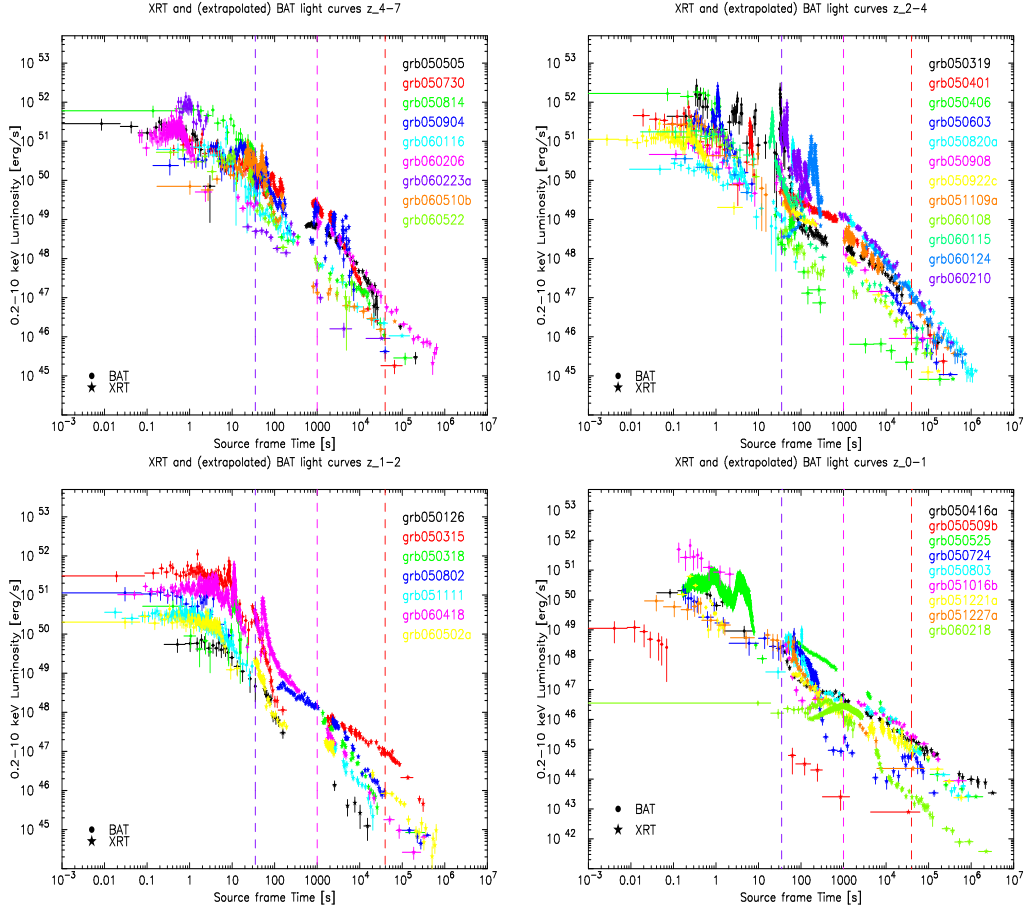

\includegraphics[height=6.5cm,width=6cm,angle=-90]{fig2a.ps}
\includegraphics[height=6.5cm,width=6cm,angle=-90]{fig2b.ps}
\includegraphics[height=6.5cm,width=6cm,angle=-90]{fig2c.ps}
\hspace{0.3cm}
\includegraphics[height=6.5cm,width=6cm,angle=-90]{fig2d.ps}
\caption{Luminosity light curves of GRBs in our sample. Top panels
show light curves of GRBs with $z\simgt4$ (left panel) and with
$2<z<simlt4$ (right panel). Bottom panels show light curves of
GRBs with $1<z<2$ (left panel) and with $z<1$ (right panel).
Vertical lines are plotted at 35 s, 1 ks and 11 hr since the trigger.}  
\label{lightcurves}
\end{figure}

In Fig. \ref{lightcurves} the light curves we computed are shown,
divided in four groups according to the redshift interval:
$z\simgt4$ (very high redshift GRBs),  $2<z\simlt4$,  $1<z<2$, and 
$z<1$ (nearby GRBs). Note that the selected sub samples have comparable
statistics. Each light curve consists of {\it i)} the
BAT light curve of the prompt emission extracted in the 15$-$150 keV 
energy band, extrapolated to flux in the 0.2$-$10 keV energy band 
using the best fit parameters of the BAT spectrum and then converted
in 0.2$-$10 keV luminosity with the K-correction required for the
best fit model of the BAT spectrum; {\it ii)} the XRT light curves,
converted to 0.2$-$10 keV luminosity with the K-correction required 
for the best fit model of the XRT spectra. 
Time dependent rate to flux and flux to luminosity conversion
have been applied when spectral variation occurred.
This lead to a very good match between BAT extrapolated
and XRT light curves.
All light curves are referred to time measured in the source rest frame
({\it i.e}. with cosmological time delay removed) starting from the trigger.

Detailed light curve fitting have been performed to determine
the continuum underlying flares during the XRT observations. 
With the only exception of GRB~050904 (that shows a light curve consisting 
of flares only, Cusumano {\it et al.} 2006), we were always able to identify 
a power law, broken power law or doubly broken power law continuum 
in agreement with the `canonical' Swift GRB light curve shape defined 
by O'Brien {\it et al.} (2006).
We found that the average time of the first break of light curves
in the source rest frame is 203 s (with 183 s of standard deviation $\sigma$), 
and average time of the second break is 20 ks ($\sigma$=35 ks).
In each panel of Fig. \ref{lightcurves} vertical lines corresponding
to 35 s, 1 ks and 11 hr since the trigger are shown. 
The selected times correspond to a typical time before the first
break of the three-phase shaped light curve (35 s), a time during the
flat intermediate phase (1 ks) and a time after the second break (11 hr).

\begin{figure}[t]
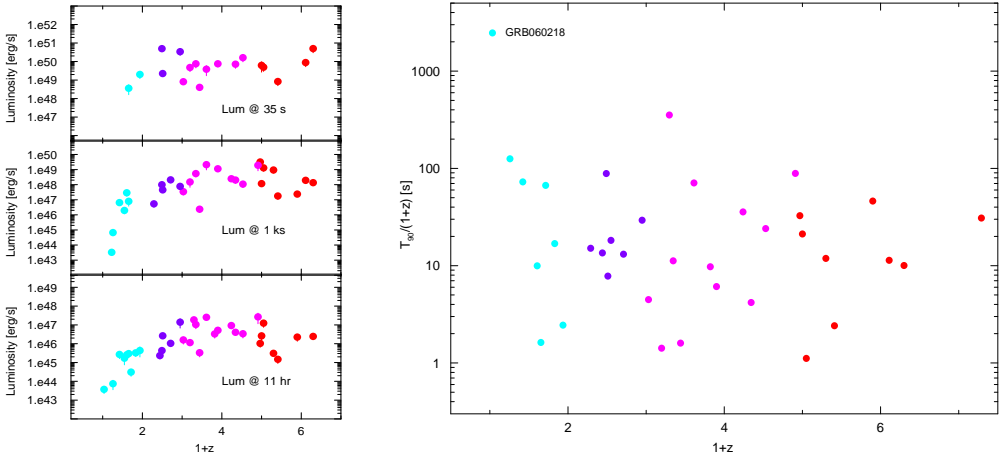

\includegraphics[height=4.5cm,width=6cm,angle=-90]{fig3a.ps}
\hspace{0.5truecm}
\includegraphics[height=8cm,width=6cm,angle=-90]{fig3b.ps}
\caption{{\bf Left panel}: luminosities of the bursts at fixed times 
(from top to bottom:35 s, 1 ks, and 11 hr) as functions of $1+z$.
{\bf Right panel}: intrinsic durations of the bursts as a function of $1+z$.}  
\label{luminosities}
\end{figure}

Luminosities of the continua at the fixed times 35 s, 1 ks and 11 hr
since the trigger have been computed in all cases when reasonable
extrapolation or interpolation of data was possible. Short bursts have
been excluded.
Results are shown in the left panel of Fig. \ref{luminosities}.
Average values and standard deviations for the four selected subsamples are 
shown in Table \ref{table}.

The right panel of Fig. \ref{luminosities} shows the intrinsic durations
of long GRBs of our sample in the source rest frame 
({\it i.e.} measured $T_{90}$ divided by (1+z)) as a function of (1+z). 
Averages of values and standard deviations for the four selected subsamples are shown in
Table \ref{table}. The average duration value for nearby bursts have
been calculated excluding GRB~060218, that would bias the result.

From these results we can see that long bursts with $z>1$ span 
approximately the same 0.2$-$10 keV luminosity range regardless of 
redshift at fixed times: around $10^{50}$ erg s$^{-1}$ after 35 s,
around $10^{48}$ erg s$^{-1}$ after 1 ks, and around $10^{46}$ erg s$^{-1}$ 
after 11 hr. Nearby long GRBs are at least one order of magnitude 
less luminous.
Moreover, intrinsic durations of long GRBs are not significantly longer 
at higher redshifts and the distribution of durations at higher redshifts
seems narrower than at lower redshifts.
The results are however very preliminary and a larger sample of GRBs
is now being investigated for selection effects.

\begin{table}[b]
\begin{tabular}{l|ll|ll|ll|ll} 
\hline
                        & \multicolumn{2}{c|}{$z<1$}  & \multicolumn{2}{c|}{$1<z<2$} & 
                          \multicolumn{2}{c|}{$2<z\simlt4$} & \multicolumn{2}{c}{$z\simgt4$} \\
                        & mean & $\sigma$ & mean & $\sigma$  & mean & $\sigma$  & mean & $\sigma$  \\
\hline
$\log L_{X}$ @ 35 s     & --   &   --  &  50.2 & 0.7    &  49.6 & 0.5 &  49.8 & 0.6  \\
$\log L_{X}$ @ 1 ks     & 46.0 & 1.5 &  47.7 & 0.6    &  48.3 & 0.9 &  48.3 & 0.8  \\
$\log L_{X}$ @ 11 hr    & 45.0 & 0.8 &  46.1 & 0.7    &  46.1 & 0.6 &  46.1 & 0.6  \\
\hline
$T_{90}/(1+z)$ (s)      & 42 & 44    &  27 & 26       &  51 & 95    &   19 & 14    \\
\hline
\end{tabular}
\caption{Average values of luminosities and durations shown in Fig. \ref{luminosities}
for the four selected redshift intervals $z<1$, $1<z<2$, $2<z\simlt4$, and $z\simgt4$.}
\label{table}
\end{table}

\acknowledgments
This work is supported at INAF by ASI grant I/R/039/04 and by COFIN MIUR grant 
prot. number 2005025417, and at Penn State by NASA contract NAS5-00136.

\vspace{1.5cm}
\appendix{Questions received:}

\begin{enumerate}
\item
Question (A. Kann): Does the sample also include HETE~II bursts with
XRT follow up redshift ({\it e.g.} GRB~050408, GRB~051022)?

Answer: No because BAT/XRT comparisons were made.

\item
Question (S. Yost): When divided into redshift bins the average $L_{X}$
are lower at low z. Is there statistical evidence that it is 
more than just a selection effect on observations?

Answer: There are selection effects at low redshift that
could lead to the observed result. But the most surprising 
feature is the flatness of $L_{X}$ at higher redshift, which is
not expected. A statistical study has not yet been done, and
will be more reliable with a larger sample, that at the present
rate of detection would be obtained in a few months.

\end{enumerate}

\end{document}